\documentclass[rmp, twocolumn]{revtex4}
\usepackage[english]{babel}
\usepackage[utf8x]{inputenc}
\usepackage{amsmath,amsfonts,amssymb,eucal,eurosym,textcomp}
\usepackage{color}
\usepackage{graphicx}
\usepackage[caption=false]{subfig}
\usepackage{pslatex}
\usepackage[colorlinks,linkcolor=red,citecolor=blue]{hyperref}
\graphicspath{{.}{ms_figures/}}

\newcommand{\xo}{\rho}

\newcommand{\Ne}{N_e}
\newcommand{\sigblock}{{\sigma_\xi}}
\newcommand{\sigblockT}{{\sigma_b}}
\newcommand{\block}{\xi}
\newcommand{\blockT}{{\xi_b}}
\newcommand{\mTtwo}{\langle T_2\rangle}

\newcommand{\EQ}[1]{Eq.~\ref{eq:#1}}

\newcommand{\FigSketch}{Fig.~1}
\newcommand{\FigTtwo}{Fig.~2}
\newcommand{\FigSFS}{Fig.~3}
\newcommand{\FigLD}{Fig.~4}
\newcommand{\FigAda}{Fig.~5}

\newcommand{\Author}{Richard A.~Neher${}^{1}$, Taylor A.~Kessinger${}^{1}$ and Boris I.~Shraiman${}^{2,3}$}
\newcommand{\Title}
{
Coalescence, genetic diversity and sexual populations under selection
}

\newcommand{\Affiliation}{${}^{1}$Max Planck Institute for Developmental
  Biology, 72076 T\"ubingen, Germany, ${}^{2}$ Kavli Institute for
  Theoretical Physics, and ${}^{3}$Department of Physics, University
  of California, Santa Barbara, 93116, CA, USA}

\begin{document}
\title{\Title}
\author{\Author}
\affiliation{\Affiliation}
\date{\today}
\begin{abstract}
\noindent
In sexual populations, selection operates neither on the whole
genome, which is repeatedly taken apart and reassembled by
recombination, nor on individual alleles that are tightly linked to
the chromosomal neighborhood. The resulting interference between linked alleles 
reduces the efficiency of selection and distorts
patterns of genetic diversity. Inference of evolutionary history from
diversity shaped by linked selection requires an understanding of these
patterns. Here, we present a simple but powerful scaling analysis
identifying the unit of selection as the genomic ``linkage block'' with
a characteristic length, $\blockT$,  determined in a
self-consistent manner by the condition that the rate of recombination
within the block is comparable to the fitness differences between
different alleles of the block. We find that an asexual model with the
strength of selection tuned to that of the linkage block provides an
excellent description of genetic diversity and the site frequency
spectra when compared to computer simulations.
This linkage block approximation is accurate for the entire
spectrum of strength of selection and is
particularly powerful in scenarios with many weakly selected loci. The
latter limit allows us to characterize coalescence, genetic diversity,
and the speed of adaptation in the infinitesimal model of
quantitative genetics.
\end{abstract}
\maketitle

In asexual populations, different genomes compete for survival, and the
fate of most new mutations depends more on the total fitness of the genome they
reside in than on their own contribution to fitness. 
As a result, beneficial mutations on one genetic background can be lost to competition
with other backgrounds, an effect known as ``clonal interference''
\cite{Gerrish:1998p5933,Desai:2007p954,neher_genetic_2013}, and likewise
deleterious mutations in very fit genomes can fix. This interference
is reduced by recombination and disappears when recombination is rapid enough such
that selection can act independently on different loci. 
Many eukaryotes recombine their genetic material by crossing over of
homologous chromosomes. As a result, distant loci evolve independently,
but nearby tightly linked loci remain coupled. Such interference,
known as Hill-Robertson interference, reduces the
efficacy of selection \cite{Hill:1966p21029,Barton:1995p3540} and
reduces levels of neutral variation.  Neutral diversity
is indeed correlated with local recombination rates in several species, suggesting that
linked selection is an important evolutionary force
\cite{Begun:1992p34015,cutter_nucleotide_2006}. One typically distinguishes 
background selection against deleterious mutations
\cite{Charlesworth:1993p36005,Hudson:1995p18197} from sweeping 
beneficial mutations, which lead to hitch-hiking
\cite{Smith:1974p34217,Gillespie:2000p28513}. Both of these processes
reduce diversity at linked loci and probably contribute to the observed
correlation \cite{hudson_how_1994}.
Another piece of evidence for the importance of linked selection comes
from the weak correlation between levels of genetic diversity and the
population size \cite{leffler_revisiting_2012}.
Whereas classic neutral models predict that diversity should increase linearly
with the population size \cite{Kingman:1982p28911}, 
in models dominated by selection the diversity depends only weakly on
the population size \cite{neher_genetic_2013}. Hence linked selection could explain 
this ``paradox of variation'' \cite{Lewontin1974}.

From the perspective of a neutral allele, any random association with
genetic backgrounds of different fitness results in fluctuations of its
allele frequency. To distinguish this source of stochasticity from
genetic drift, Gillespie coined the term ``genetic draft''
\cite{Gillespie:2000p28513}. While genetic draft is well understood when
caused by strongly selected mutations whose dynamics is deterministic
at high frequencies
\cite{Walczak:2011p45228,weissman_limits_2012,Barton:1995p3540}, the
cumulative effect of many weak effect mutations has mainly been
addressed using simulations \cite{McVean:2000p19278,Gordo:2002p36599}.
Many populations harbor substantial
heritable phenotypic variation which in an unknown way depends on a
large number of polymorphisms in the genome. The majority of these
polymorphisms are likely to have small effects on phenotypes and
fitness. Collectively, however, they still can dominate phenotypic
variation \cite{Yang:2010p35988} and possibly fitness variation.
This limit is known as the infinitesimal model in quantitative genetics. 
Quantitative genetics, however, typically ignores linkage between loci 
and the maintainance of genetic diversity
\cite{Bulmer_1980,Lynch:1998p8721}. 

Here, we characterize the structure of genealogies, genetic diversity
and the rate of adaptation in sexual populations in the limit of
numerous weakly selected alleles.  We build on recent progress
in our understanding of genealogies in adapting asexual populations
\cite{neher_genealogies_2013,desai_genetic_2013,Brunet:2007p18866} and we
will first review these results briefly. We
will then present a scaling argument that reduces the problem of
coalescence within an sexually reproducing population to an asexual 
population with suitably scaled parameters. This correspondence
allows us to predict levels of genetic diversity, coalescence time
scales, and site frequency spectra. Our results hold regardless of
whether the polymorphisms originated as weakly deleterious or
beneficial mutations and thus cover weak effect background selection or
adapation. We confirm the validity of the mapping to the asexual model
by comparing its predictions with numerical simulations of evolving
sexual populations. We use this approximation to demonstrate that in the
limit of numerous weakly selected mutations, the rate of adaptation
scales as the square root of recombination rate.

\section{Results}
In asexual populations all loci share the same genealogical history, and
the fate of a lineage depends on the fitness of the entire genome.
If fitness depends on a large number of polymorphic loci
with comparable effects, the fitness distribution in the population will
be roughly Gaussian, and the fittest individuals are $x_c \approx \sigma
\sqrt{2\log N\sigma}$ ahead of the fitness mean, where $\sigma^2$ is the
total fitness variance in the
population \cite{Tsimring:1996p19688,Rouzine:2003p33590,Desai:2007p954}. 
In large asexual populations, only individuals in the high
fitness nose have an appreciable chance to contribute to future generations. 
It will take those individuals roughly $\sigma^{-1}\sqrt{2\log N\sigma}$
generations to dominate the population. Hence the probability
that two randomly chosen individuals had a common ancestor
$\sigma^{-1}\sqrt{2\log N\sigma}$ generations ago is of order one, i.e.,
their ancestral lineages have likely coalesced. A more thorough analysis
of coalescence in adapting asexual populations can be found in 
Refs.~\cite{neher_genealogies_2013,desai_genetic_2013} \footnote{In
  Ref.~\cite{neher_genealogies_2013} it is shown that $\mTtwo \approx
  \sigma^{2}/D$. Since $\sigma^2 \approx (24D^{2} \log N\sigma)^{1/3}$, $\mTtwo \approx c\sigma^{-1} \sqrt{2\log N\sigma}$ with a $c=\sqrt{12}$.}. 
In small populations with $N\sigma \ll 1$,
coalescence is dominated by neutral processes (non-heritable
fluctuations in offspring number known as genetic drift). The average 
number of generations back to the most recent common ancestor of any
pair of extant genomes, a.k.a.~the pair coalescence time, is given by:
\begin{equation}
  \label{eq:asex}
  \mTtwo \approx
  \begin{cases}
    N & N\sigma \ll 1 \\
    c\sigma^{-1}\sqrt{2\log N\sigma} & N\sigma\gg 1
  \end{cases}
\end{equation}
where $c$ is a constant of order one that captures deviations from Gaussianity
that depend on details of the model. For the infinitesimal model studied here $c=\sqrt{12}$. 

In an attempt to extend applicability of the neutral coalescent, one sometimes defines an
``effective population size'', $\Ne$, equal to $\mTtwo$ regardless of whether
coalescence is neutral or not \cite{Charlesworth:2009p33963}. By definition a neutral model with
$\Ne=\mTtwo$ predicts the same levels of genetic diversity, but the
statistical properties of the genealogies dominated by selection are quite
different and cannot be papered over simply by redefining the population
size. We will therefore avoid the term $\Ne$ and stick to $\mTtwo$. 
For the approximately neutral case, $N\sigma\ll
1$, the coalescent tree is of the Kingman type
\cite{Kingman:1982p28911}. As $N\sigma$ increases,
coalescence is more and more driven by the amplification of fit genomes,
which generates a very skewed offspring number distribution over
timescales of order $\sigma^{-1}$. As a result, the genealogies resemble the
Bolthausen-Sznitman coalescent (BSC)
\cite{Bolthausen:1998p47390,Brunet:2007p18866} with very different
statistical properties. Two representative
coalescent trees sampled from asexual populations, one neutral and one
rapidly adapting, are shown in \FigSketch A. 

\begin{figure}[th]
  \centering
  \includegraphics[width=0.98\columnwidth]{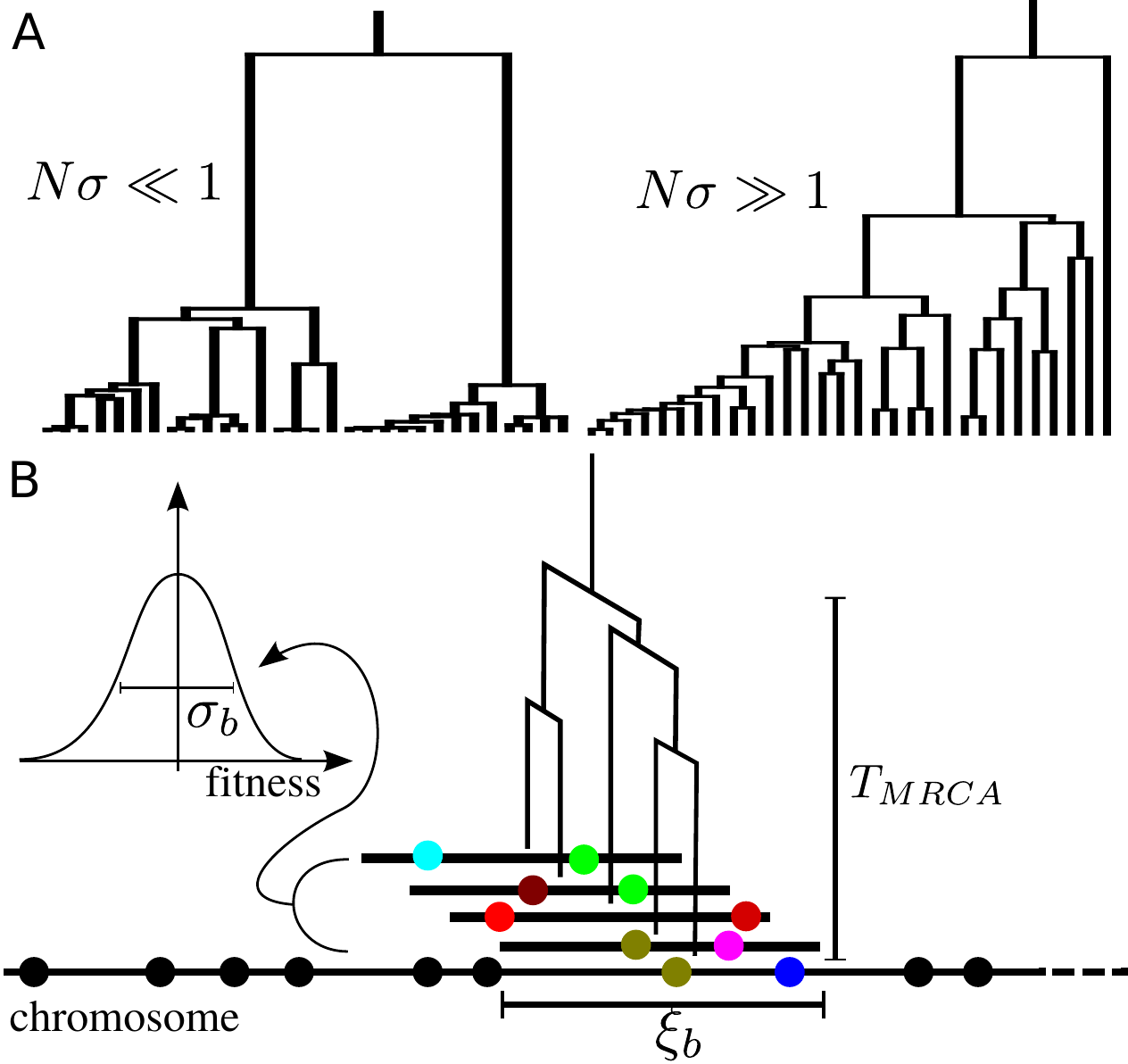}
  \caption{
    Coalescence in neutral and adapting populations. Panel (A) shows a
    typical coalescent tree from neutral and adapting asexual
    populations (left and right, respectively). In adapting populations,
    coalescent trees branch asymmetrically and contain
    approximate multiple mergers. Panel (B) illustrates 
    asexual blocks in sexual populations. The sketch depicts a
    representative chromosome at the bottom with polymorphisms
    indicated as balls. Different loci within segments shorter than
    $\blockT$ share most of their genealogical history, i.e., have
    trees similar to the one indicated in the center of the
    segment. Coalescence within this segment of length $\blockT$ is
    either neutral or driven by the fitness differences between
    different haplotypes spanning these segments. 
    The fitness distribution of these haplotype blocks is indicated as
    inset. Distant parts of the chromosome are in linkage
    equilibrium, and the tree changes as one
    moves along the chromosome. The succession of changing trees
    is the ancestral recombination graph.  
  }
  \label{fig:sketch}
\end{figure}

\subsection*{Sexual populations and recombination.} In contrast
to asexual evolution,  recombination decouples different loci in sexual
populations -- the
further apart, the more rapidly. The typical length of the segment that
is not interrupted over a time $t$ along one ancestral lineage decreases
with time as
\begin{equation}
  \label{eq:block_length}
  \block = \frac{L}{1+L\xo t} \approx \frac{1}{\xo t}
\end{equation}
where $\xo$ is the crossover rate and $L$ is the length of the chromosome.
The second approximation is justified whenever $\block\ll L$. If polymorphisms
affecting fitness are spread evenly across the genome and are dense (the
infinitesimal model), we expect that
different segregating haplotypes in a region of length $\block(t)$ 
harbor fitness variation proportional to the segment length 
\begin{equation}
  \label{eq:sigma_block}
  \sigblock^2 = \frac{\block(t)}{L}\sigma^2 \ .
\end{equation}
This fitness variance shrinks with time as the block length decreases.
While initial fitness differences between blocks are large, they
are chopped into smaller blocks so rapidly that selection has no time to
amplify the fittest of these early large blocks. But the rate at which
blocks are chopped up decreases as they get shorter, and at some point
the rate of chopping them up is outweighed by the amplification of the
fittest blocks by selection. The latter happens when fitness differences
between haplotypes of this block are comparable to the recombination rate.
More precisely, the relevant block length $\block(t)$ is the
length that survives over the time scale of coalescence, i.e.,
$\blockT=\block(\mTtwo)$. In large enough populations, the time scale of
coalescence itself is determined by these fitness differences via
\EQ{asex}. In constrast to asexual populations, only the
fitness variance, $\sigblockT^2$,  within the linkage block of length $\blockT$ is
relevant rather than the total variance $\sigma^2$ (see illustration in \FigSketch{}B).
Using $\mTtwo = c{\sigblockT}^{-1}\sqrt{2\log
N\sigblockT}$ in \EQ{block_length}, we find for the length of linked
blocks 
\begin{equation}
  \label{eq:blockTc}
  \blockT = \frac{\sigblockT}{c\xo \sqrt{2\log N\sigblockT}} \ .
\end{equation}
LD measured in populations samples should decay over this length scale.
Substituting $\blockT$ into \EQ{sigma_block} yields
\begin{equation}
  \label{eq:self_consist}
  \sigblockT = \frac{\sigma^2}{L\xo\, c\sqrt{2\log N\sigblockT}} \quad
  \mathrm{and} \quad \blockT = \frac{\sigma^2}{2L\xo^2\, c \log
    N\sigblockT} \ .
\end{equation}
Hence the time scale of coalescence and neutral diversity are given by
the inverse of the fitness variance per map length $R=L\rho$ with a
logarithmic correction (see also
\cite{Santiago:1998p34629,Hudson:1995p18197} for the case of strongly selected mutations). 
To arrive at this result, we have assumed that coalescence is driven
by selection, i.e., we have assumed $N\sigblockT\gg 1$. If this
condition is not satified, local coalescence will be approximately
neutral. In this case $\mTtwo = N$ and the LD extends over
$\blockT\sim (N\rho)^{-1}$ nucleotides. Empirically, we observe a smooth and rapid
crossover between these two regimes (see below and \FigTtwo).

The condition for draft dominance, $N\sigblockT\gg 1$, is more
stringent in sexual populations than in asexual populations, in which it is
$N\sigma\gg 1$. In other words, recombination reduces interference and
results in drift dominated coalescence over a larger parameter range. 
We predict now that the results for genetic diversity in the
asexual coalescent apply with $\sigblockT^2$ as the local fitness variance
and that linkage disequilibrium between common loci extends over
a distance $\blockT$. We will validate these predictions by forward
simulations of different population models.

\subsection*{Constant selection in the infinitesimal model.}
We first consider a model of a population whose fitness
variance is set by external (environmental) factors in which the selected
trait depends on many weak effect polymorphisms and de novo mutations; see Model and
Methods. This model might be a first approximation to scenarios where
selection pressures are dictated by a changing environment, an evolving
immune system, or a breeder who imposes a constant 
 artificial selection. We simulate our population using a
discrete generation model with an approximately constant population size
and a finite number of sites in the genome as implemented in FFPopSim
\cite{zanini_ffpopsim:_2012} (see Methods).  We track the genealogy of a locus in the center of
the chromosome, which allows us to study properties of representative
coalescent trees. 

After allowing the population to equilibrate, 
we sample the evolving population in roughly
$\mTtwo$ intervals and measure $T_2$, $T_{MRCA}$, the site frequency spectrum (SFS), and the
linkage disequilibrium (LD) between polymorphisms at intermediate
frequencies ($[0.1, 0.9]$). We perform these simulations for many
combinations of parameters. For each of these combinations, we calculate
$\sigblockT$ according to \EQ{self_consist}. \FigTtwo~shows
that the average pair coalescence time $\mTtwo$ approaches $N$ for
$N\sigblockT\to 0$ and that it is proportional to $\sigblockT^{-1}$ (with
logarithmic corrections) for $N\sigblockT\gg 1$ as predicted.

\begin{figure}
  \centering
  \includegraphics[width=0.91\columnwidth]{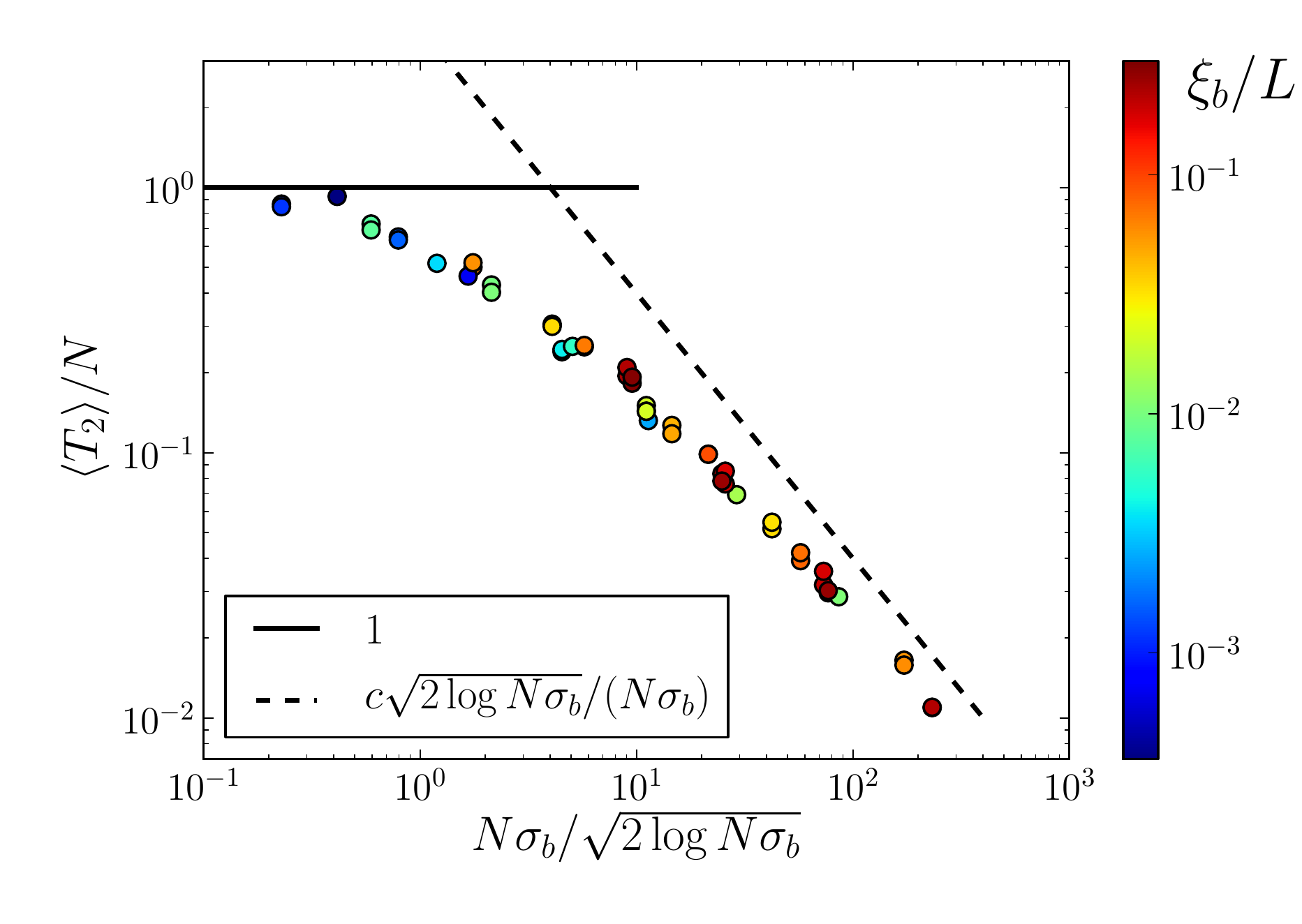}
  \caption{Coalescence in sexual populations. The figure shows the
    average pair
    coalescence time $\mTtwo$ relative to the neutral expectation as a
    function of $N\sigblockT$ determined using \EQ{self_consist}. 
    For $N\sigblockT\ll 1$, $\mTtwo\approx N$,
    while  $\mTtwo = c{\sigblockT}^{-1}\sqrt{2\log
  N\sigblockT}$ otherwise. }
  \label{fig:Ttwo}
\end{figure}

In addition to a reduction in genetic diversity, we predict that the
local genealogies will resemble samples from the BSC
rather than the Kingman coalescent whenever $N\sigblockT\gg 1$. \FigSFS~shows a
collection of SFS colored by the  $N\sigblockT$. With increasing $N\sigma_b$, the SFS 
smoothly interpolate between the expectations for the Kingman coalescent and the BSC. 
As soon as the SFS starts deviating from
the prediction of the Kingman coalescent, Tajima's D and related
measures turn negative. For large $N\sigma_b$, we find a non-monotonic SFS with a steep 
divergence $f(\nu)\sim\nu^{-2}$ for rare alleles characteristic of the BSC.

\begin{figure}
  \centering
  \includegraphics[width=0.91\columnwidth]{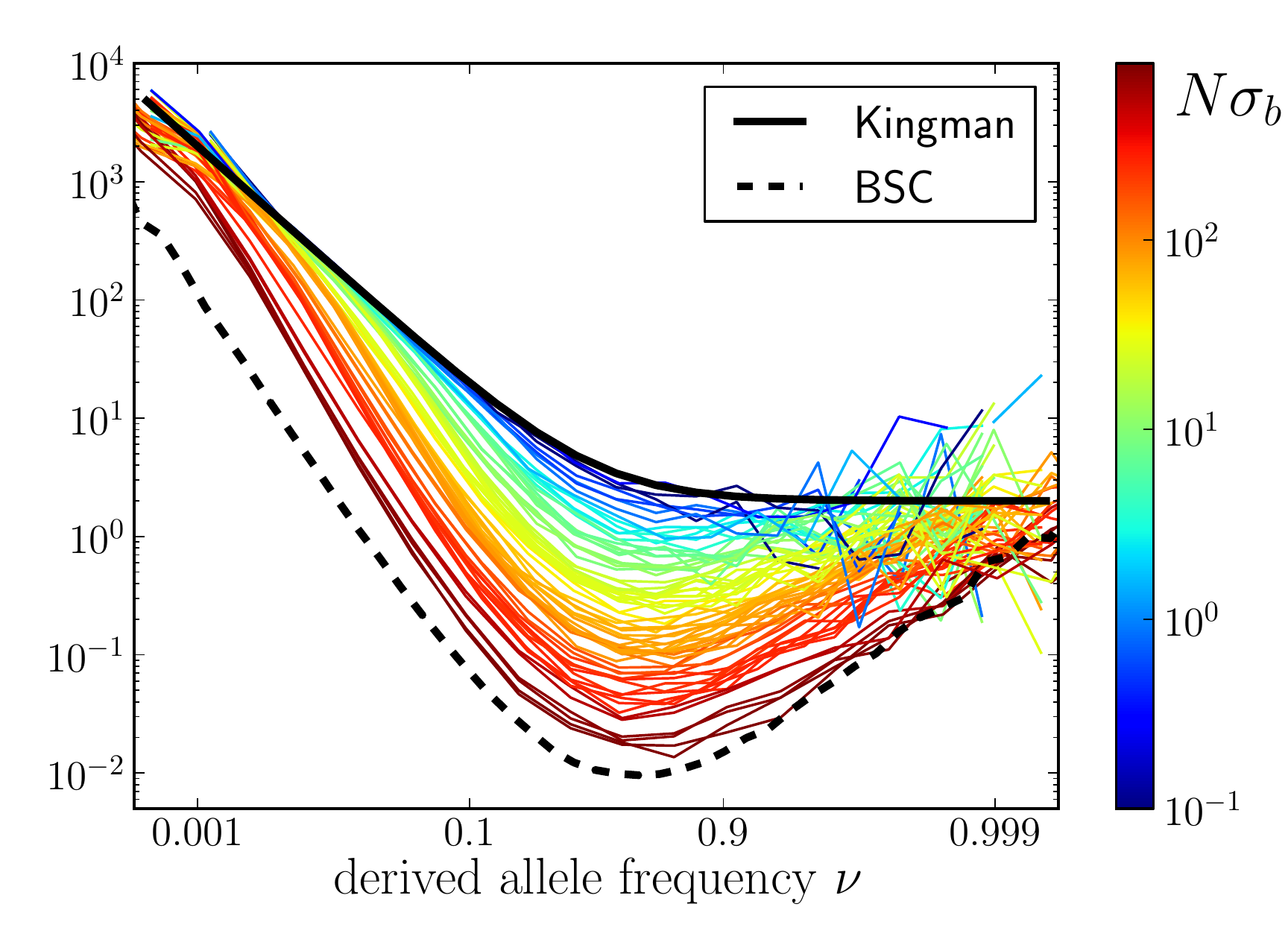}  
  \caption{Site frequency spectra (SFS). The figure shows the SFS, 
    normalized by $\Theta = 2N\mu$, for a large number of parameter combinations. 
    Color indicates the value of $N\sigblockT$. For large $N\sigblockT$, the SFS display the
    non-monotonicity characteristic of the BSC (dashed line), while the
    SFS are well described by the prediction from Kingman's coalescent
    (solid line) if $N\sigblockT\ll 1$. The BSC curve serves as a guide to the
    eye since its normalization depends on $N\sigblockT$.}
  \label{fig:SFS}
\end{figure}

Another important feature of diversity in sexual populations is
the genomic distance across which loci share much of their
genealogy. This can be quantified by measuring the correlations between
loci (LD) at different distances. In order for our picture to be
consistent, the extent of LD should be approximately equal to $\blockT = (\xo
\mTtwo)^{-1}$. We measured LD as $r^2(d)$ for different distances $d$ and
plot its distance dependence against $d/\blockT$; see \FigLD. As
predicted, the distance over which loci are correlated
is well described by $\blockT =(\xo \mTtwo)^{-1}$.

\begin{figure}
  \centering
 \includegraphics[width=0.91\columnwidth]{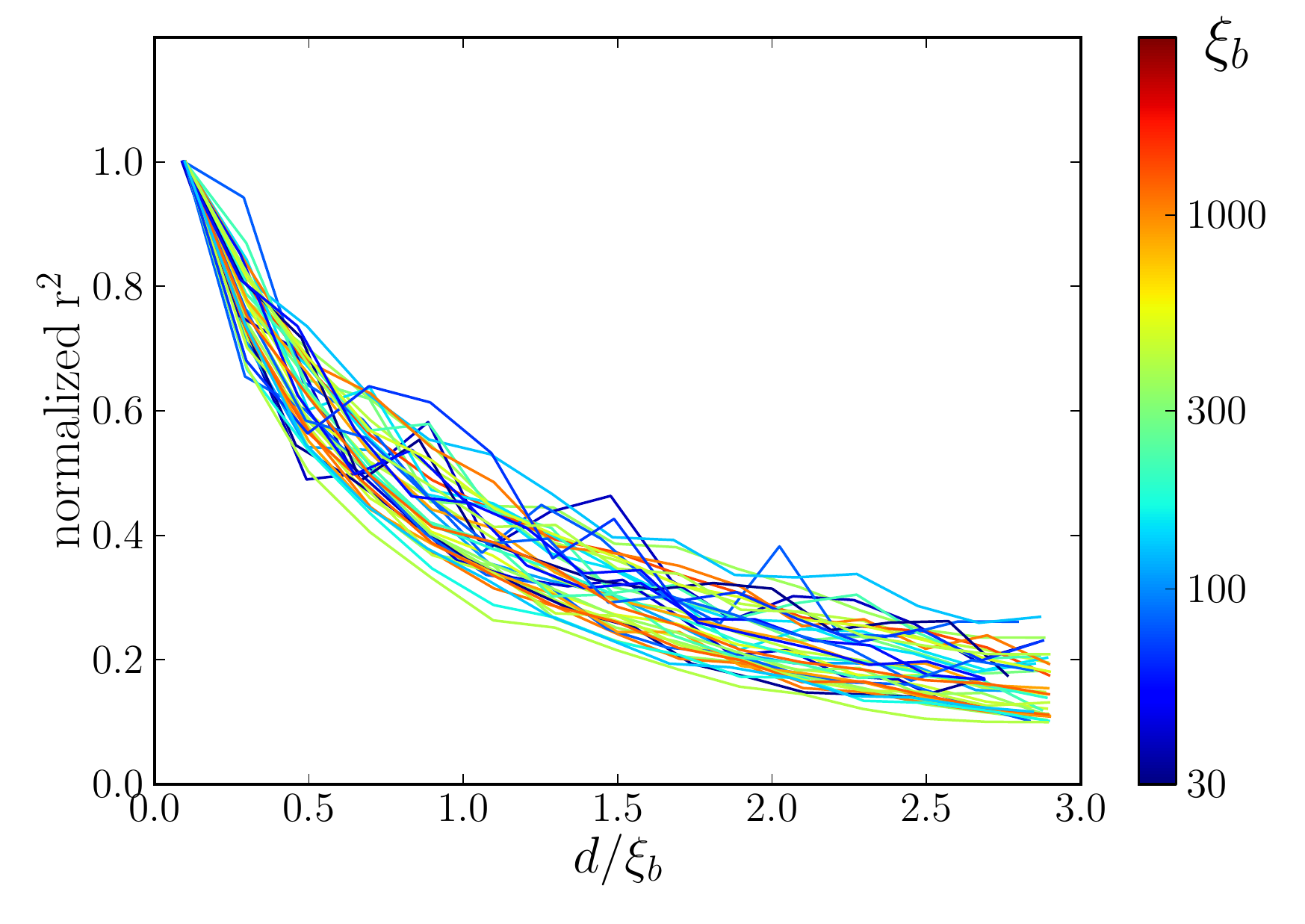}
  \caption{Correlation length along the genome. The figure shows
    linkage disequilibrium, quantified as average $r^2$, between pairs of loci at different
    distances (the curves are normalized to their value at zero
    distance). The x-axis shows the distance between loci $d$ rescaled by
    $\blockT$ determined using \EQ{block_length} with $t$ equal to the measured pair
    coalescence time. After this rescaling, the distance dependence of all
    simulations follow approximately the same master curve, which
    shows that LD extends for $\approx \blockT$. }
  \label{fig:LD}
\end{figure}

\subsection*{Frequent small effect mutations.}
In the model studied above, fitness variance was set by external factors. 
We now consider a model where the fitness variance and diversity are set
by a balance between frequent novel mutations of small effect and the removal of variation by
selection, i.e., fixation or loss of  alleles. 
This type of model has been studied for asexual populations
\cite{Tsimring:1996p19688,Cohen:2005p45154}. Using these results, we expect that 
the fitness variance within a block of length $\blockT$ is given by
\begin{equation}
  \label{eq:sigma_total_diffusive}
  \sigblockT^2 \approx \frac{\blockT \mu \langle s^2 \rangle}{2} \mTtwo \ .
\end{equation}
Here, $\mu$ is the mutation rate, and $\langle s^2 \rangle$ is the second
moment of the distribution of mutational effects. Note than in this infinitesimal limit
it is irrelevant whether mutations are deleterious or beneficial -- only the second
moment of the fitness effect distribution is important.
The quantity $D=\frac{\blockT\mu \langle s^2
  \rangle}{2}$ is the ``diffusion'' constant of haplotype fitness in the
absence of selection. \EQ{sigma_total_diffusive} implies that fitness variation accumulates
over the time it takes a few lineages to dominate the population, which is
approximately given by the half the pair coalescence
time\cite{neher_genealogies_2013}. 
Substituting \EQ{block_length} with $t=\mTtwo$ into \EQ{sigma_total_diffusive}, we find
\begin{equation}
  \label{eq:sigma_block_DM}
  \sigblockT^2 =  {\frac{\mu\langle s^2 \rangle}{2\xo}}
\end{equation}
Remarkably, this variance of the effectively asexual blocks is simply
the ratio of variance injection per nucleotide, $\mu \langle s^2 \rangle$, and the
crossover rate (at least while $N\sigblockT\gg 1$). The coalescence time
cancels! We therefore find for $\mTtwo$ 
\begin{equation}
  \label{eq:Tc_diffusive}
  \mTtwo \approx
  \begin{cases}
    N & N\sqrt{\mu\langle s^2 \rangle\xo^{-1}} \ll 1 \\
   c\sqrt{\frac{\xo\log N\sigblockT}{\mu \langle s^2\rangle} } & N\sqrt{\mu\langle s^2 \rangle\xo^{-1}} \gg 1\\ 
  \end{cases}
\end{equation}
where $c$ is again a constant of order 1. In the limit where coalescence
is driven by selection, the total rate of adaptation is therefore 
\begin{equation}
  \label{eq:adaptation_diffusive}
  \sigma^2 \approx cL\sqrt{\xo \mu \langle s^2\rangle \log N\sigblockT} \ .
\end{equation}

These results apply to steadily adapting populations (i.e.~scenarios
where beneficial mutations dominate), populations suffering from a
mutational meltdown, or populations where the two processes balance. 
We simulate the lattermost using a model with recurrent mutations such that the population settles into
a dynamic equilibrium where the fixation of beneficial mutations is roughly canceled out
by that of deleterious mutations \cite{Goyal:2012p47382}. The predictions for neutral
diversity, LD and the SFS match the simulation results very
well. Supplementary
figure S1 shows plots analogous to \FigTtwo~through 4.
The prediction for the total fitness variance,
\EQ{adaptation_diffusive}, is compared to the simulation results in
\FigAda. We investigated additional models to demonstrate the robustness of the
conclusions regarding model assumptions and simulation method.
Supplementary figure S2 shows neutral diversity, LD, and SFS for a model
in which unique beneficial mutations are injected at sites that become
monomorphic. Supplementary figure S3 shows results for a bona fide
infinite sites model of chromosomes of length $1$ that undergo one
crossover per generation and accumulate beneficial or deleterious
mutations at rate $U$. 
In all of these cases, the observed diversity agrees well with the predictions of
\EQ{Tc_diffusive} and the SFS show the expected crossover from the
Kingman to the BSC predictions as $N\sigblockT$ increases. 

\begin{figure}
  \centering
  \includegraphics[width=0.95\columnwidth]{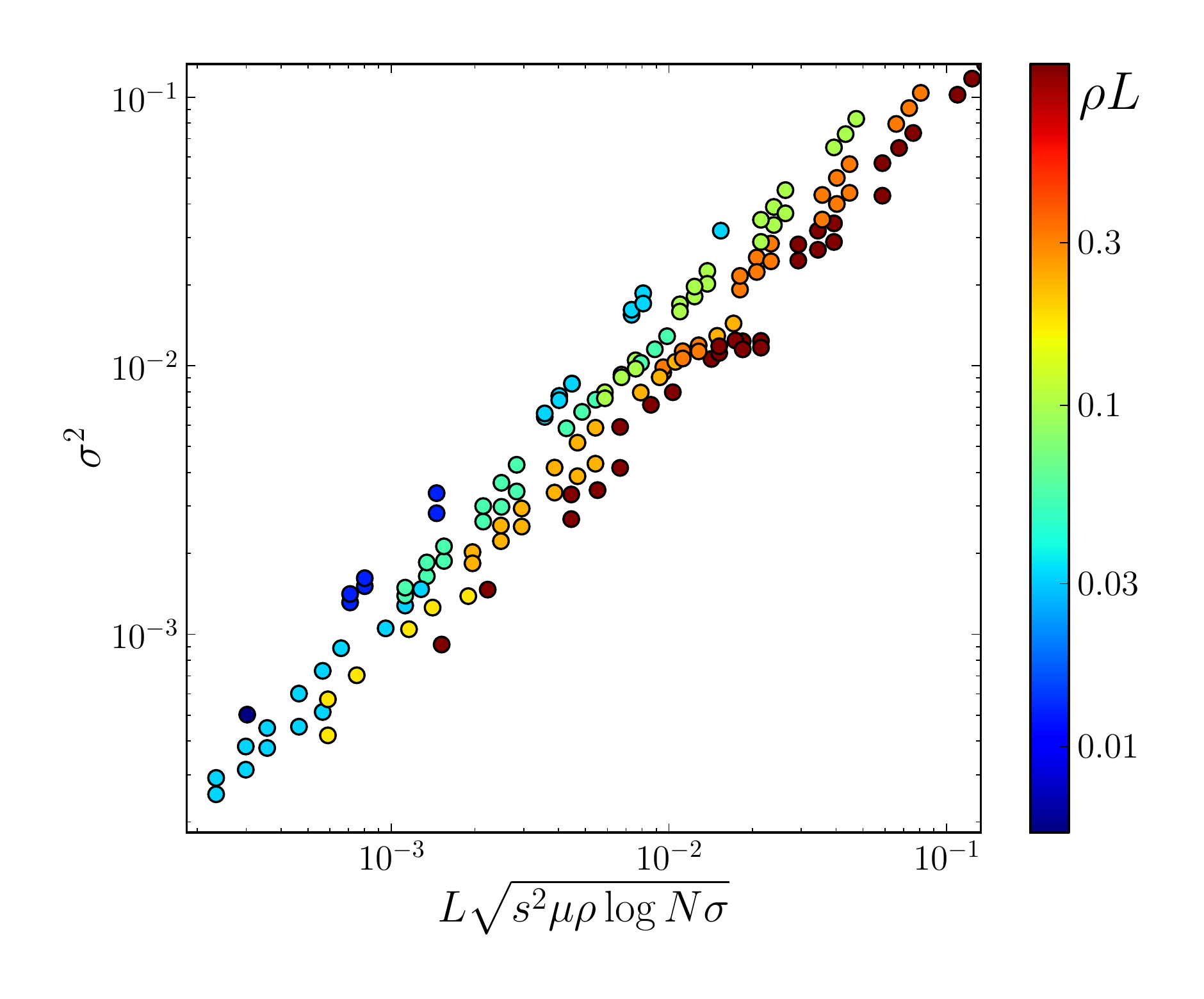}
  \caption{The total fitness variation due to frequent weak effect mutations in a model
    where deleterious and beneficial mutations balance each other. The
    color shows the average number of crossovers per simulated
    segment. There is a residual dependence on $\xo$ due to
    large corrections to the asymptotic behavior.}
  \label{fig:adaptation_dynbalance}
\end{figure}

\subsection*{Loosely linked loci.}
Our analysis has focused on the effect of fitness variation in
short effectively asexual blocks. As discussed above, the total
strength of selection $\sigma$ can be much larger than the fitness
differences within effectively asexual blocks $\sigblockT$. However,
a particular locus only remains linked to distant polymorphisms
for a short time, and the contribution of these distant loci
averages out. For our focus on the effect of tightly linked loci to be
valid, the integral contribution of such loosely linked loci 
to drift and draft should be small compared to the effect of
fitness variation $\sigblockT$ within the segment.
Loosely linked loci are amenable to a
perturbative analysis known as Quasi-Linkage Equilibrium
\cite{Kimura:1965p3008,Neher:2011p45096}. In
Ref.~\cite{Neher:2011p45096} it is shown that the stochastic dynamics of
the allele frequency $\nu_i$ at locus $i$ due to loosely linked loci is
described by the following Langevin equation:
\begin{equation}
  \label{eq:allele_langevin}
  \frac{d}{dt}\nu_i(t) = \nu_i(1-\nu_i) s_i +2\mu(1-2\nu_i) + \sum_{i\neq
    j}D_{ij}(t)s_j + \eta_i(t) \ ,
\end{equation}
where $D_{ij}(t)$ is the LD between loci $i$ and $j$, $s_j$ is the
fitness effect of the derived allele at locus $j$, and $\eta_i$ is
random noise with autocorrelation function $\langle
\eta_i(t)\eta_i(t')\rangle =N^{-1}\delta(t-t')$, representing genetic
drift. If the two loci are loosely linked, i.e., the
crossover rate $c_{ij}$ between them is much larger than the effect of selection on either
of them, $D_{ij}$ is also a fluctuating quantity. 
The autocorrelation function of $D_{ij}$ is \cite{Neher:2011p45096}
\begin{equation}
  \label{eq:LD_autocorr}
  \langle D_{ij}(t)D_{ij}(t')\rangle = \frac{\nu_i(1-\nu_i)\nu_j(1-\nu_j)e^{-c_{ij}|t-t'|}}{2Nc_{ij}} \ .
\end{equation}
Given this autocorrelation, we can now integrate over fluctuations due
to genetic drift and loosely linked selected loci to obtain a
renormalized diffusion coefficient, i.e., the reduction of the
``effective population size''. Reproducing Eq.~44 of
\cite{Neher:2011p45096}, we have
\begin{equation}
  \label{eq:Ne}
  \frac{N}{N_e} = 1+\frac{1}{2}\sum_{i\neq j} \nu_j(1-\nu_j)\frac{s_j^2}{c_{ij}^2}
\end{equation}
This result is similar to results in
\cite{Nordborg:1996p18149,Hudson:1995p18197,Santiago:1998p34629}
in that it shows that the level of drift is increased by a factor that
depends on the square of the ratio of selection and linkage, averaged
over the genome.

If we now consider the integral effect of all loci further away than
$\xi$, it is always dominated by the loci at the smallest distance, so
that $N/\Ne -1 \sim ( \sigma / R )^2 ( \xi / L)^{-1} $ (obtained as
a continuum approximation to the sum in \EQ{Ne}, $R=\rho L$). Hence,
provided that $\xi/L > ( \sigma / R)^2 $ -- a condition that obtains when
fitness variation at distant loci is sufficiently small or the loci are sufficiently distant --
their effect can be accounted for by a simple rescaling of the effective
population size \cite{weissman_limits_2012}; this is the {\it weak
  draft} regime. Note, however, that the recombination rate between
distant loci is ultimately limited by the outcrossing rate and that
distant loci can have substantial effects in facultatively sexual
populations \cite{Neher:2010p30641,weissman_limits_2012}.

The negligible effect of loosely linked loci is a consequence of two
types of averaging that are apparent in \EQ{LD_autocorr}: (i) The 
associations between these distant loci are transient and average out
over time. This manifests itself in the decay time of $c_{ij}^{-1}$ in
\EQ{LD_autocorr}. (ii) Different individuals carry different alleles at
these distant loci, and hence their fitness effect is averaged over
different descendents. As a consequence, the auto-correlation in \EQ{LD_autocorr} is
proportional to $(Nc_{ij})^{-1}$. Together, these two averages result
in the $1/c_{ij}^2$ contribution of loosely linked loci.

For the more tightly linked loci, i.e., $\xi <  \xi_*=( \sigma / R)^2 L
$, the behavior crosses over
to the {\it strong draft} regime. This crossover length scale $\xi_*$ is 
controlled entirely by the {\it local}
quantities: the recombination rate per base pair $\rho$ and the local
fitness variance density. Furthermore, $\xi_*$ is in general larger than
$\blockT$, with 
$\xi_*/\blockT \sim \log (N \sigblockT )$. This ratio corresponds to the
reduction in the block size during the span of time between local
selection effects first coming into play and the coalescence time. In
the limit of $\log (N \sigblockT ) \gg 1$ recombination events within
the $\xi_*$ block must be reckoned with, but for more realistic population sizes,
we have shown above that focusing on the $\blockT$-sized asexual segment
captures the effects of strong draft quite well.

\subsection*{Length distribution of segments identical by descent (IBD).}
The structure of genealogies has implications for the length $\ell$ of
IBD segments in pairs of individuals. Their distribution, $p(\ell )$, is
directly related to the distribution of pair coalescence times, $q(T_2
)$, via the relation $p(\ell ) \sim \int dT_2 q(T_2 ) e^{-\rho \ell T_2}$. In
neutrally evolving populations of constant size, pair coalescence times
are exponentially distributed with mean $\mTtwo=N$. Consequently, the
length of IBD segments is distributed as $p(\ell) \sim 1/(1+\rho \ell
\mTtwo)$ and has a long slowly decaying tail.  If $N\sigblockT\gg 1$,
coalescence is accelerated on average but predominantly happens after lineages have reached the upper
tail of the fitness distribution of different alleles of a
linkage block. Hence the distribution of pair coalescence times is
peaked at $\mTtwo$ rather than being exponential; comp.~Fig.~3 in 
ref.~\cite{neher_genealogies_2013}. This shift in the distribution of $T_2$
with relatively rare very recent coalescence has the consequence that $p(\ell ) \sim
e^{-\rho \ell \mTtwo}$ is approximately exponential. Long IBD
segments are therefore much less likely than in the neutral case with the same $\mTtwo$.


\section{Discussion}
In most sexual populations, the histories of different chromosomes or
loci far apart on a chromosome are weakly correlated. Nearby loci,
however, are more tightly linked, which results in correlated histories
and linkage disequilibrium. Since the density of heterozygous sites
is $\pi = 2\mu \mTtwo$ and the length scale of LD is $\blockT = (\xo
\mTtwo)^{-1}$, the typical number of SNPs in one linkage block is 
 $n \approx \mu/\xo$. If $n$ is much larger than one, and
a sizeable fraction of those SNPs affect fitness, different haplotypes
segregating within such a block will display a broad distribution in
local fitness with a variance that we have denoted by
$\sigblockT^2$. Neutral alleles linked to haplotypes drawn from this
distribution will be affected by linked selection. This in turn 
results in genealogies different from standard
neutral models but similar to the Bolthausen-Sznitman coalescent (BSC)
characteristic of rapidly adapting asexual populations
\cite{neher_genealogies_2013,Neher:2011p42539}.

In regions of high recombination in obligately outcrossing species the 
number of polymorphisms per linkage block, $n$, is of order one and linked selection
will mainly result from the occasional strong selective sweep \cite{Sella:2009p26729}. But recombination 
rates vary by orders of magnitude across the genome \cite{comeron_many_2012}
and $n\gg 1$ in low recombination regions. In those regions, the cumulative effect
of many weakly selected polymorphisms is expected to be important. 
This holds in particular for species that outcross rarely, such as many plants, 
nematodes, yeasts, and viruses
\cite{bomblies_local-scale_2010,barriere_high_2005,Neher:2010p32691,tsai_population_2008}.
This type of linked selection will overwhelm genetic drift if $N \sigblockT >1$. The fitness
variance per block is given by $ \sigblockT^2 = \langle s^2\rangle \pi
\blockT$, where $\langle s^2\rangle$ is the second moment of the effect
distribution of polymorphisms. Hence
we require $N^2\langle s^2\rangle > (\pi \blockT)^{-1} = n^{-1}$.  Provided
$n$ is large enough, even nominally neutral ($Ns <1$) polymorphisms collectively
dominate the dynamics of haplotypes of length $\blockT$. In this
infinitesimal limit, the nature of linked selection is irrelevant and
our results apply to any mix of deleterious and beneficial mutations as long
as the effects of individual mutations are weak and their number is large. 

\subsection*{Relation to previous work.}
Most previous work on genetic draft and selective interference
considered mutations with strong effects that behave deterministically
at high frequencies, whereas we focus on weak effect mutations.
Reduction of genetic diversity by sweeping beneficial mutations was first
discussed by Maynard Smith \cite{Smith:1974p34217}; see also
\cite{Barton:1998p28270,Gillespie:2000p28513,Kaplan:1989p34931,Braverman:1995p34932}. 
In these models, genetic
diversity is determined by the typical waiting time between two
successive selective sweeps close enough to affect a given locus.
Similarly, deleterious mutations reduce diversity at linked sites. 
Assuming that mutations have a large detrimental effect on fitness and
happen with rate $\mu$ per site, it was shown in refs.~\cite{Hudson:1995p18197,Nordborg:1996p18149}
that the reduction of genetic diversity is a function of $\mu/\xo$. As in our
analysis here, the strongest effect on genetic diversity comes from
tightly linked loci. Our analysis of loosly linked loci is similar to
the work by Santiago and Caballero \cite{Santiago:1998p34629}. The latter,
however, breaks down at tight linkage, and the crossover to the asexual 
behavior is essential for a consistent description in the limit of many
weakly selected loci. This limit has mainly been studied using computer 
simulations \cite{McVean:2000p19278,gordo_mullers_2002,messer_frequent_2013}, and few analytical
results are available. 

Weissman and Barton \cite{weissman_limits_2012} investigated the rate of adaptation and
its effect on diversity using scaling arguments similar to the one
presented here. In their model, adaptation is driven by individual
selective sweeps. The duration of a sweep explicitly sets the
time scale $\mTtwo$ on which coalescence happens. In this model, the
speed of adaptation is proportional to the map length. In
contrast, our model assumes many weak effect mutations, and the
time scale of coalescence is set by $\sigblockT$, which is
self-consistently determined and itself depends on model parameters
such as $\xo$ and $\mu \langle s^2\rangle$. We can
recover their result for the rate of adaptation by
setting  $\mTtwo\sim s^{-1}$ and $\blockT
\sim s/\xo$. With these assumptions, we obtain
\begin{equation}
  \label{eq:weissman}
  \sigma^2 \sim L \xo s
\end{equation}
instead of \EQ{adaptation_diffusive}. The model used in 
Ref.~\cite{weissman_limits_2012} applies to a limit where at most one
strongly selected and sweeping mutation falls into one linkage block,
but our analysis considers the opposite limit. The basic properties of
genealogies and SFS are expected to be qualitatively
similar in the limit of one sweep per block. If the contribution from
weak mutations is negligible while sweeps are common, the coalescence
properties will be dominated by sweeps at different distances. This
limit has been studied in \cite{Durrett:2005p40919} and also results in
a multi-merger coalescent.

Other types of models are appropriate if the rate of outcrossing is small compared 
to the standard deviation in fitness \cite{Rouzine:2005p17398,Neher:2010p30641,Neher:2011p42539} 
or if recombination proceeds via horizontal transfer of short pieces of DNA
\cite{Cohen:2005p5007,Neher:2010p30641}. In these cases, one finds 
a very strong dependence of the rate of adaptation on the rate of outcrossing
or horizontal transfer. Rare recombination has the potential to dramatically increase
fitness variance because many loci are in strong LD.

In summary, we have characterized the effect of dense weakly selected polymorphisms
on genetic diversity, which might be the source of much of the phenotypic variability
we observe \cite{Yang:2010p35988,Lynch:1998p8721}. Our analysis provides a consistent
genealogical framework for the infinitesimal model of quantitative genetics. 
This limit of weakly selected mutations has so far eluded analytical understanding.
We derived equations that relate the mutational input and the rate of recombination
to neutral diversity and the site frequency spectra. Because 
genetic diversity (neutral or not) is directly accessible 
in population resequencing experiments, our results should be of practical
relevance when interpreting such data. 
Furthermore, one is often interested in identifying particular mutations
that arose in response to specific environmental challenges. If
successful, those mutations tend to be of large effect and fall outside
the scope of our model. Importantly, strong 
adaptations only perturb a fraction of the genome (more precisely a
segment of length $\approx s(\xo \log Ns)^{-1}$, where $s$ is the
selection coefficient). Our model provides the background on
top of which such singular adaptations can be sought, and understanding
the statistical patterns of diversity and linkage within this null model
is essential for reliable inference. 

\section{Acknowledgements}
We would like to thank Fabio Zanini for stimulating discussions and help
with FFPopSim and Guy Sella for very useful comments on the manuscript. 
This work is supported by the ERC starting grant HIVEVO
260686 to R.A.N and in part by the NSF PHY11-25915 grant to
KITP. B.I.S. acknowledges support from NIH R01 GM086793.

\section{Methods}
We use a model with discrete generations, haploid individuals, an approximately constant
population size, and a finite number of sites in the genome, as
implemented in FFPopSim \cite{zanini_ffpopsim:_2012}. We simulate a
fraction of a chromosome of length $L$, where outcrossing happens with
rate $\xo$ between randomly chosen gametes and results in a single crossover. 
If $\xo L\ll 1$, no 
recombination happens in most cases. In addition to forward simulation,
we also track the genealogy of a central locus, which allows us to measure pair
coalescence times, the time to the MRCA, and the neutral SFS directly
(this functionality is implemented in a more recent release of FFPopSim;
see \url{http://code.google.com/p/ffpopsim}). For all parameters, we
produce equilibrated populations by simulating for 10
$T_{MRCA}$. Subsequent measurements of population parameters start from
these equilibrated populations and sample the population roughly twice
every $\mTtwo$ as estimated from our theoretical arguments. All scripts
associated with this paper can be obtained from
\url{http://git.tuebingen.mpg.de/reccoal}.

\subsection{Constant selection}
To maintain a constant fitness variance $\sigma^2$, we rescale
the selection coefficients associated with individual loci each
generation accordingly. Mutations are introduced into a random individual whenever a locus
becomes monomorphic, i.e., the previously introduced mutation is lost
or has fixed (see \cite{Neher:2011p42539}). This allows us to simulate
a large number of sites efficiently in a limit where the overall
mutation rate is small compared to $\mTtwo$. In this way, we keep all $L$ loci
polymorphic without employing a high mutation rate, which would result
in frequent recurrent mutations.
We simulate a grid of parameters with $N$ taking the values $[1000,
3000, 10000]$, $\sigma$ the values $[0.01, 0.03, 0.1]$, and $L\rho$ five
logarithmically spaced values between $0.1\sigma$ and $1.0 \sigma$. For the
analysis, simulations were filtered so that $\blockT>30$ and
$\blockT<L/3$. To prevent invalid logarithms, $\log(N\sigblockT)$ was
replaced by $\log(N\sigblockT+2)$ in \EQ{self_consist}.

\subsection{Dynamic Balance}
In this set of simulations, we simulate a genome consisting of finite sites in a
constant fitness landscape where mutations at each locus have a small
effect $s$. Mutations are injected at random with rate $\mu$ at each
locus. In contrast to the models above, where mutations are injected
only when a locus is monomorphic, we allow recurrent and back
mutation to make the dynamic balance state possible.
The grid of parameters used was $L\in [3000, 10000]$, $N\in
[1000,3000, 10000]$, $s \in [-0.001, -0.003, -0.01]$, $L\mu \in [1,3,
10, 30]$, and $L\rho$ logarithmically spaced between $s$ and $1.0$. 
For the analysis, simulations were filtered such that $\blockT>30$, 
$\blockT<L/3$, and $\mTtwo \mu <0.5$.

\bibliography{/ebio/ag-neher/share/bibliography/bib}
\clearpage
\setcounter{figure}{0}

\onecolumngrid
\section*{Supplementary figure 1: recurrent mutations with weak effects}

\begin{figure}[h]
  \centering
  \includegraphics[width=0.9\columnwidth]{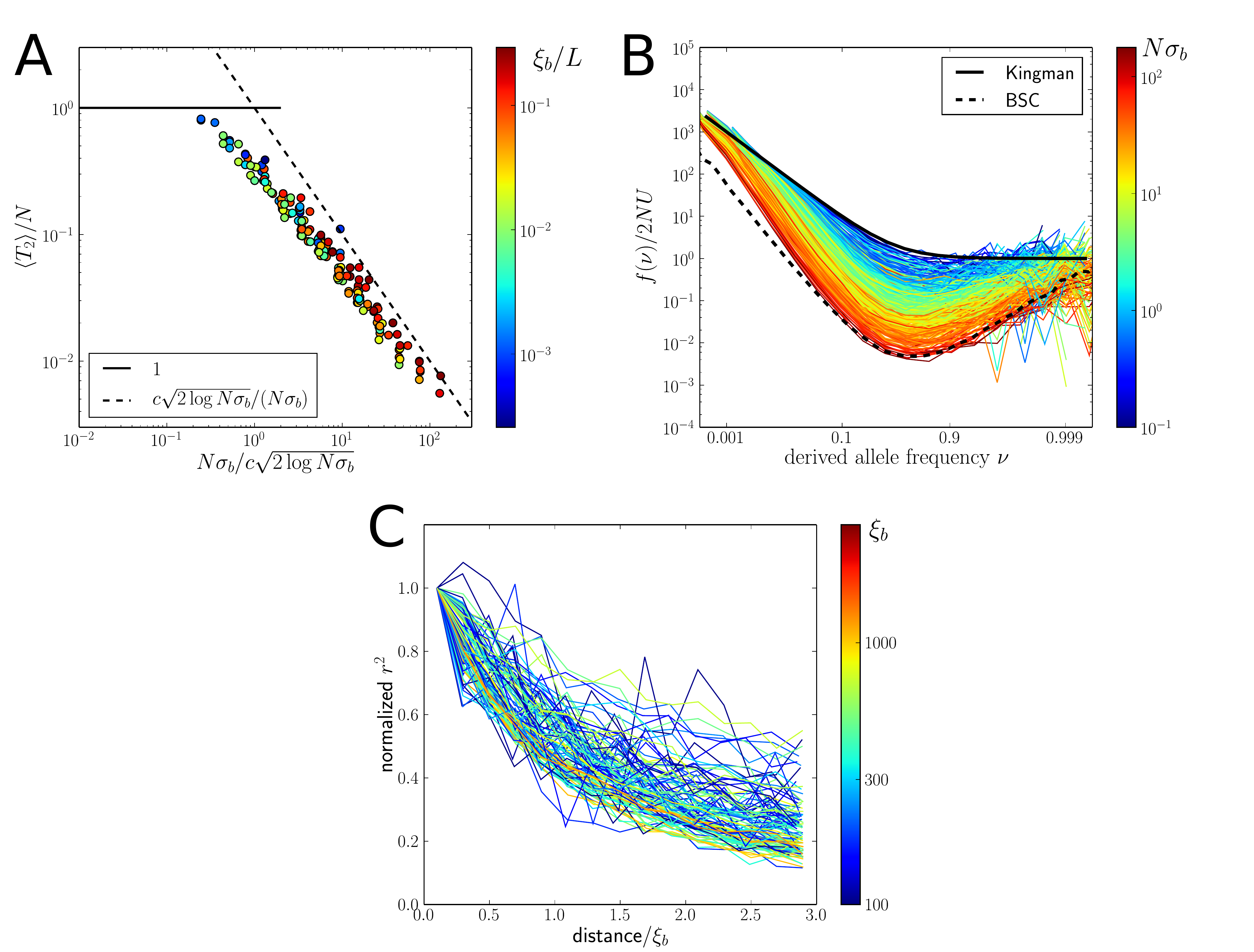}
  \caption{Genetic diversity in populations with recurring mutations between a preferred and unpreferred state with weak effect. Panel A shows the pairwise coalescence time compared to the analytical predictions in the limits of large and small $N\sigblockT$. Panel B shows the SFS normalized to $\Theta = 2N\mu$ (the SFS are obtained from local coalescent trees). Different curves are colored by their respective $N\sigblockT$ values. The BSC curve serves as a guide to the eye since its proper normalization depends on $N\sigblockT$.  Panel C shows the decay of LD measured as $r^2$ and normalized with its value at short distances. The $x$-axis is rescaled by $\blockT$. The resulting collapse demonstrates that LD extends over distances $\blockT$. 
The grid of parameters used for simulations was $L\in [3000, 10000]$, $N\in
[1000,3000, 10000]$, $s \in [-0.001, -0.003, -0.01]$, $L\mu \in [1,3,
10, 30]$, and $L\xo$ logarithmically spaced between $s$ and $1.0$. 
For the analysis, simulations were filtered such that $\blockT>30$, 
$\blockT<L/3$, and $\mTtwo \mu <0.5$}
  \label{fig:dark_matter_suppfig}
\end{figure}

\clearpage
\section*{Supplementary figure 2: beneficial mutations with fixed effect}

\begin{figure}[h]
  \centering
  \includegraphics[width=0.9\columnwidth]{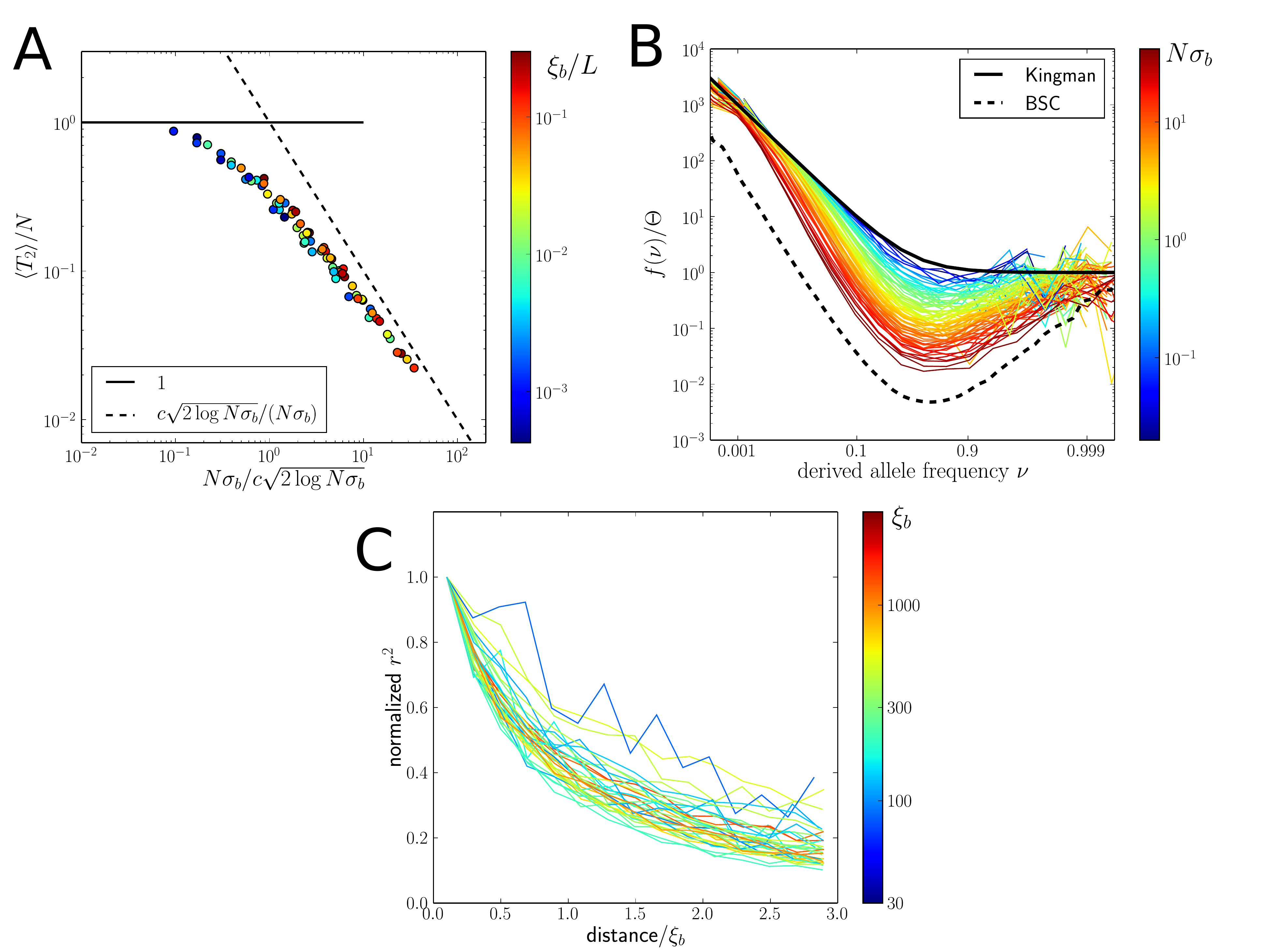}
  \caption{Genetic diversity in populations with frequently sweeping beneficial mutations. Panel A shows the pairwise coalescence time compared to the analytical predictions in the limits of large and small $N\sigblockT$. Panel B shows the SFS normalized to $\Theta = 2N\mu$ (the SFS are obtained from local coalescent trees). Different curves are colored by their respective $N\sigblockT$ values. The BSC curve serves as a guide to the eye since its proper normalization depends on $N\sigblockT$.  Panel C shows the decay of LD measured as $r^2$ and normalized with its value at short distances. The $x$-axis is rescaled by $\blockT$. The resulting collapse demonstrates that LD extends over distances $\blockT$. 
In these simulations, mutations are introduced into a random individual
whenever a locus becomes monomorphic, analogous to the simulations with
constant fitness variance discussed in the main text. However, in this
set of simulations, the fitness variance is a fluctuating quantity. 
The grid of parameters used was $L\in [3000, 10000]$, $N\in
[1000,3000, 10000]$, $s \in [0.001, 0.003, 0.01]$, and $L\xo$ logarithmically spaced between $s$ and $1.0$. For the analysis, simulations were filtered such that $\blockT>30$ and
$\blockT<L/3$. 
}
\label{fig:fixed_effect}
\end{figure}

\clearpage
\section*{Supplementary figure 3: deleterious and beneficial mutations in an infinite sites model}

\begin{figure}[h]
  \includegraphics[width=0.9\columnwidth]{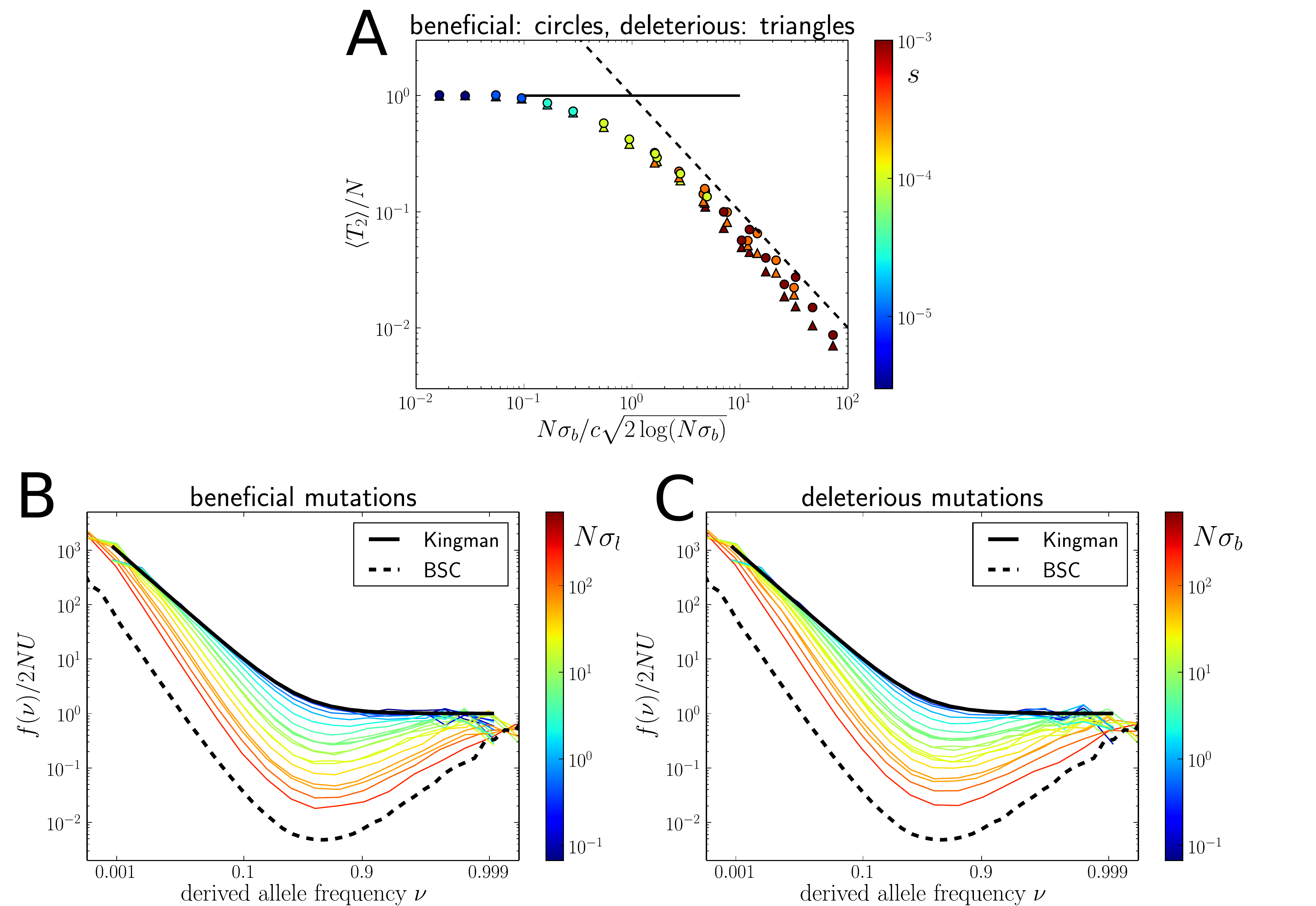}
   \centering
   \caption{Beneficial and deleterious mutations in a bona fide infinite sites model. Panel A shows the pairwise neutral diversity or coalesence time for simulations with beneficial (circles) and deleterious (triangles) mutations. The color of the symbols indicates the absolute effect size of mutations. Panels B\&C show the corresponding SFS for beneficial and deleterious mutations, respectively. The SFS are obtained from histograms of the frequency of neutral polymorphisms and normalized to $\Theta = 2NU_{n}$, where $U_n$ is the total neutral mutation rate.
These results are obtained with a model that assumes chromosomes of length $1$ 
that undergo exactly one crossover per generation. The chromosomes mutate at random places 
in the interval $[0,1]$. With probability $0.5$, mutations are neutral; otherwise they have an
effect $s$ on fitness. We simulate a total mutation rate $U\in [10,30,100]$ with effect sizes $[3\times 10^{-5}, 10^{-4}, 3\times 10^{-3}, 10^{-3}, 3\times 10^{-3}]$ (positive and negative) for population sizes $N\in [1000, 3000, 10000]$. The SFS and the neutral diversity follow the predictions of the analysis presented in the paper. LD was not investigated using this model.}
\label{fig:infinite_sites}
\end{figure}

\end{document}